\documentclass[aps,prd,twocolumn,groupedaddress,showpacs]{revtex4}
\usepackage{graphicx}
\usepackage{dcolumn}
\usepackage{bm}

\begin{document}

\title{Tetraquark particle in the string model}

\author{Masaharu Iwasaki}
\email{miwasaki@cc.kochi-u.ac.jp}
\author{Takahiko Fukutome}
\affiliation{Department of Physics, Kochi University, Kochi 780-8520, Japan}

\date{\today}

\begin{abstract}
We discuss possibility of the existence of tetraquark states made of four quarks in the string (flux tube) model. The new particle is composed of a diquark and an anti-diquark which are connected by a color flux. It is shown that the vibrational and rotational excited states of the string explain some non-$\bar{q}q$ mesons observed experimentally. Moreover we discuss the decay widths of such tetraquarks with the use of the Schwinger mechanism.

\end{abstract}

\pacs{12.39.Ki, 12.40.Yx, 13.25.-k, 13.30.Eg}

\maketitle

\section{Introduction}

It is well known that there are only two kinds of hadrons in Nature: mesons and baryons. The former are made of a quark and an antiquark and the latter made of three quarks. Their existence is explained by quantum chromodynamics (QCD) \cite{W80} that is the fundamental theory for hadron physics. There is, however, a possibility of the existence of other type hadrons, such as four-quark particles. In fact a possible four-quark state has been suggested experimentally by the several groups \cite{A01}.

In this paper we discuss the possible existence of a four-quark state which we call a ``tetraquark"; it is made up of two quarks and two antiquarks. Since Jaffe predicted such hadrons with the use of the MIT bag model \cite{J77a,J77b}, many non-$\bar{q}q$ mesons have been discussed by many authors \cite{R04}-\cite{MPPR05}. Particularly it is conjectured that the nonet of light scalar mesons [$f_{0}$(600), $f_{0}$(980), $a_{0}$(980), $\kappa$(800?)] with masses below 1 GeV seem to fit a $qq\bar{q}\bar{q}$ description. Instead a heavier scalar nonet above 1 GeV seems to be largely $\bar{q}q$ configuration. The light scalars seem to have a significant component of $qq\bar{q}\bar{q}$ and the pairs of quarks (and antiquarks) are separately correlated in a flavor, color and spin antisymmetric state. A diquark and an anti-diquark neatly couple together to give a nonet of color singlet scalar mesons. There is also evidence for strong diquark correlations in the baryon spectrum \cite{APE93}.

We will consider a possibility that this strong diquark correlation leads to a molecular-like configuration, $(qq)-(\bar{q}\bar{q})$ with the use of the chemical bond-like expression. To this end, we will study the four-quark system by using the hadron string model or the flux tube model. According to the model, mesons are made up of a quark and an antiquark which are connected by the classical gluon field \cite{IP83}. The baryons are also made up of a quark and a diquark connected by the same flux \cite{MKB89,M90}. Using this simple model, F.Takagi and one of the present authors (M.I.) have shown that many excited states of hadron observed experimentally can be identified as the vibrational and rotational excited states of this one-dimensional string \cite{IT99,INST03}.

Since any hadron must be a color-singlet, the only possible four-quark state is a $qq\bar{q}\bar{q}$ state. There are here two configurations: (a) $(\bar{q}q)-(\bar{q}q)$ and (b) $(qq)-(\bar{q}\bar{q})$. We suppose that other configurations $(q)-(q\bar{q}\bar{q})$ and $(\bar{q})-(qq\bar{q})$ are not so stable. Assuming the one-gluon exchange interaction, the $q\bar{q}$ system is attractive in the color singlet state (meson) so that the system (a) probably decays into two mesons at once. On the other hand $qq$ ($\bar{q}\bar{q}$) system is attractive in the color anti-triplet (3bar) state, which means that it is equivalent to $\bar{q}$ ($q$) in the color SU(3) space. That is the system (b) is the same as the meson as far as the color degree of freedom. Therefore a four-quark state considered in this paper is assumed to be composed of a diquark and an anti-diquark connected by the same gluon flux as the meson.

It should be emphasized here that our approach is different from that in Jaffe's papers. According to them, the nonet of light scalar is described by the $s$-wave $qq\bar{q}\bar{q}$ sector in the MIT bag model; all the quarks are in the same $s$-state in the bag potential. On the other hand, a scalar in our model is described by the $s$-sate of the diquark and anti-diquark where each quark is also $s$-state. The latter $s$-state is quite different from the former one. The wave function of the quark would be a superposition of many states of the higher angular momenta in the MIT bag potential. The relation of our model to the MIT model is analogous to that of the cluster model to the shell model in nuclear structure models. Therefore the tetraquark discussed in this paper are not the nonet of light scalar but other heavier mesons.

In the next section (Sec.{\rm II}), we present our flux tube model and derive the mass spectra and decay widths using the WKB approximation and the Schwinger mechanism. The numerical calculations of these quantities are carried out in the Sec.{\rm III}. Finally (Sec.{\rm IV}) several discussions and summary are given.

\section{A flux tube model}

According to \cite{IT99}, our system is described by the relative distance $r$ between the diquark and the anti-diquark and the rotation angle $\theta$ of the string. The geometry is shown in Fig.1 where their masses is denoted by $M_{1}$ and $M_{2}$. 
 \begin{figure}
 \includegraphics[width=\linewidth]{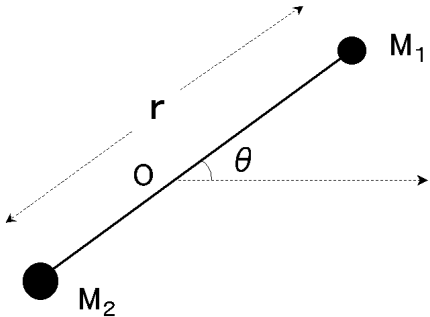}
 \caption{The schematic picture of the hadron string.}
 \end{figure}
We take the origin as the center of mass. Since our discussion is restricted to the nonflavored mesons, we assume $M_{1}=M_{2}\equiv M$. Then the Lagrangian of this system is given by
\begin{equation}
L(r,\dot{r},\omega)=-2M\sqrt{1-v^2}- V(r,\omega), \label{eq:lagrangian}
\end{equation}
where $v^2=(\dot{r}^2+r^2\omega^2)/4$ is the square velocity of each diquark. The first term represents the action of the diquark and the anti-diquark and the second one the potential energy of the system which is represented by
\begin{equation}
V(r,\omega)=2\int_0^{r/2} a\sqrt{1-r^2\omega^2} dr -\frac{e}{r}.
\end{equation}
The first term denotes the energy of the gluon flux with the string tension $a$ (the energy per unit length) \cite{JT75} and $\sqrt{1-r^2\omega^2}$ comes from the Lorentz contraction of the transversal motion. The second one represents the one-gluon exchange potential with the coupling constant $e$. This expression for the potential is known as Cornel potential and confirmed by the Lattice QCD simulation \cite{B01,AK05}. It should be noted that this Lagrangian is also valid for the $\bar{q}q$ system (ordinary mesons) because their color charges are the same as those of the $qq-\bar{q}\bar{q}$ one.

The Hamiltonian of our system is defined by $H=\dot{r}p+\omega l-L$ where $p=\partial L/\partial \dot{r}$ and $l=\partial L/\partial \dot{\theta}$. Then it is written as
\begin{equation}
H(p,l,r)=\frac{2M}{\sqrt{1-v^2}}+\int_0^{r/2}\frac{2a}{\sqrt{1-r^2\omega^2}}dr-\frac{e}{r}. \label{eq:hamiltonian}
\end{equation}
where the factor $\sqrt{1-r^2\omega^2}$ in the right-hand side represents the Lorentz contraction along the transverse direction of the string. The excited states of the system are described by the Schr\"{o}dinger equation, $H(p,l,r)\psi(r,\theta)=E\psi(r,\theta)$. Carrying on the separation of variables $\psi=(\chi/r)\exp{im\theta}$, this equation is reduced to
\begin{equation}
H\left(-i\frac{\partial}{\partial r},m,r\right)\chi(r)=E\chi(r), \label{eq:Schroedinger}
\end{equation}
where $l$ is the angular momentum quantized as $m=0,1,2,3,\cdots$. This equation can be solved using the WKB approximation and the excited energy $E$ is obtained by the Bohr-Sommerfeld formula,
\begin{equation}
\oint p_{r}dr=(n+1/2)h. \;\;\;(n=0,1,2,3,\cdots)
\end{equation}
The right hand side should be replaced by $(n+3/4)h$ when $m=0$. The full derivation of the formula is explained in Ref. \cite{IT99}.

Here let us discuss the qualitative properties of the mass spectra (neglect the masses and the one-gluon exchange term for convenience). In the case of $m=0$ (vibrational motion), the energy becomes $E=2p+ar\equiv ar_{0}$ ($r_{0}$ is the maximum length). This leads to $\oint pdr=(1/2)ar_{0}^{2}\approx 2\pi n$. Thus we have
\begin{equation}
E^2=4\pi a n.
\end{equation}
In the case of $n=0$ (rotational motion), the energy is $E=2m/r+ar$ if the kinetic energy is approximated by the centrifugal potential. This is minimized by $r=\sqrt{2m/a}$ so that $E=2ar$. These relations lead to
\begin{equation}
E^2=8a m.
\end{equation}
In both cases, it is concluded that the energy squared $E^2$ is proportional to the quantum numbers $n$ or $m$ and the ratio of the proportional constants is approximately $\pi:2$.

Next let us consider the decay width of our system. The possible decay process that we consider is a pair production of $\bar{q}q$ inside the flux tube which is known as Schwinger mechanism \cite{S51}-\cite{GM83}. The probability of the pair production in a unit space-time volume in the tube is given by
\begin{equation}
w=\frac{a^2}{4\pi^3} \sum_q \sum_{n=1}^\infty \frac{1}{n^2}\exp{\left(-\frac{n\pi {m_q}^2}{a}\right)}.
\end{equation}
where $a$ is the string tension and $\sum_q$ indicates a summation over all quark flavors with mass $m_q$ (q=u,d,s) \cite{G79}-\cite{GM83}. The probability of the decay (pair production) in the time interval $dt$ is given by
\begin{equation}
dW=\int \mid \chi(r) \mid^2 wv(r)drdt,
\end{equation}
where $v(r)$ denotes the volume of the string with the length $r$ and is given by
\begin{equation}
v(r)=2\sigma \int_{0}^{r/2}\sqrt{1-r^2\omega^2}dr.
\end{equation}
In the right-hand side, $\sigma$ denotes the area of the cross section of the tube in the rotating coordinates system. The decay width of our system is given by\begin{equation}
\Gamma=\frac{dW}{dt}=\int |\chi(r)|^2 w v(r)dr,\label{eq:decay}
\end{equation}
This formula gives the decay widths of the excited hadrons due to the quark pair production. As is shown in this formula, the decay width is proportional to the length of the string. This means that it increases as the excited energy increases.

\section{Numerical results}

We are in a position to carry out the numerical calculations of our tetraquark particles. Our string model has two components, the color flux and the diquark. The former is characterized by the string tension $a$ and the cross sectional area $\sigma$ and the latter is concerned to the diquark mass. The coupling constant $e$ will be used to be $e=0.4$, which is a usual value \cite{B01}. The parameters concerning to the color flux should be taken as the same values as those of $\bar{q}-q$ system, because a diquark (anti-diquark) has the same color charge as that of an antiquark (quark). These parameters have been determined so as to reproduce the experimental mass and decay width of $\rho(770)$.

When we take $a=0.11{\rm GeV^{2}}$ and ($M_{1}=M_{2}=10$MeV), the calculated vabrational spectrum of the rho meson is shown in Fig.2. 
 \begin{figure}
 \includegraphics[width=\linewidth]{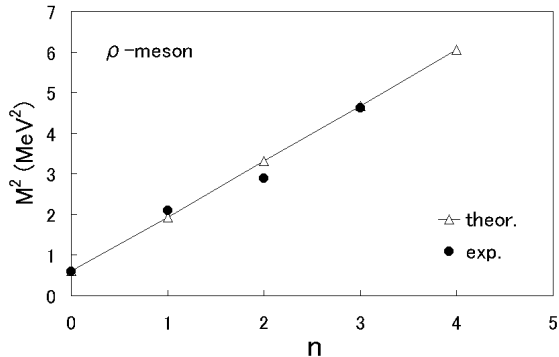}
 \caption{The vibrational excited states of the rho meson. The quantum number of the vibrational motion is denoted by $n$.}
 \end{figure}
In the case of $m=0$, the interval of the integral (5) is $0<r<r_{0}$ so that the integral diverges. Therefore the cutoff parameter $\epsilon=0.01{\rm fm}$ $(\epsilon<r<r_{0})$ has been introduced so as to reproduce the mass of $\rho(770)$. Its decay width ($\Gamma=150$MeV) determines the other parameter, $\sigma=75{\rm GeV^{-2}}$. We use these parameters in later calculations hereafter.

As for the diquark mass, we take the most simple diquark which is composed of $u$ and $d$ quarks. In other words we restrict ourselves to the flavor SU(2). In this case, the most stable diquark is color anti-triplet, flavor-singlet and spin-singlet taking into account the Fermi statistics. Since the nucleon is made up with this diquark and the remaining quark, our diquark mass is determined by the nucleon mass N(939) as is done in Ref. \cite{IT99}: $M=300$MeV.

First it is adequate to discuss the possible quantum numbers of our tetraquarks, which has configuration $(qq)-(\bar{q}\bar{q})$. Since the diquark (anti-diquark) is flavor-singlet and spin-singlet, it is isoscalar ($I=0$) and has not spin angular momentum. The total angular momentum $J$ is nothing but the orbital one: $J=m$. The G-parity operation defined by $G=Ce^{i\pi I_{2}}$ is the same as that of the ordinary parity $P$, because the charge conjugation $C$ is equivalent to $P$ in our system. Therefore we should seek mesons with $I^{G}(J^{PC})=0^{\pm}(J^{\pm\pm})$.

Let us discuss the vibrational excited states with the angular momentum $m=0$. The square of mass is drawn as a function of the quantum number $n$ in the Fig.3. 
 \begin{figure}
 \includegraphics[width=\linewidth]{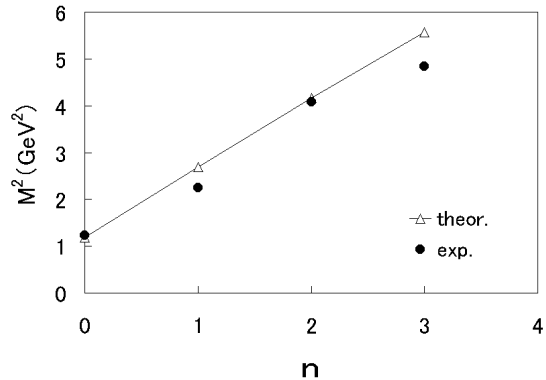}
 \caption{The vibrational excited states of the tetraquark. The quantum number of the vibrational motion is denoted by $n$.}
 \end{figure}
The closed circles in the figure represent the corresponding particles with $0^{+}(0^{++})$ tentatively. They are $f_{}$(1110), $f_{0}$(1500), $f_{0}$(2020), $f_{0}$(2200) \cite{PDG}. The head state $f_{}$(1110) is $0^{+}({\rm even}^{++})$ and its spin $J$ is not determined except for $J=$even integer. It is seen our simple model reproduces such particles fairly well. The trajectory is an almost straight line as is shown in the Eq.(6). 

Next let us consider the rotational excited states with $n=0$, which is called Regge trajectory. In the Fig.4, the square of mass is shown as a function of the angular momentum. 
 \begin{figure}
 \includegraphics[width=\linewidth]{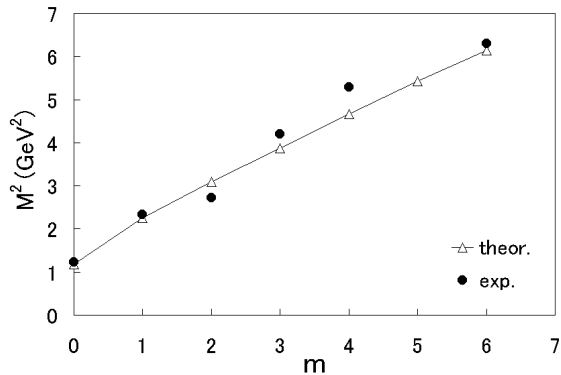}
 \caption{The rotational excited states of the tetraquark. The orbital angular momentum is denoted by $m$}
 \end{figure}
As discussed above, the quantum numbers of the Regge trajectory should be $0^{+}(0^{++})$, $0^{-}(1^{--})$, $0^{+}(2^{++})$, $0^{-}(3^{--})$, $\cdots$. So the closed circles represent the corresponding particles tentatively: $f_{}$(1110), $\omega_{1}$(1420), $f_{2}$(1640), $\omega_{3}$(1945), $f_{4}$(2050), $\omega_{5}$?, $f_{6}$(2510). The candidate for $\omega_{5}$ has not been observed yet. We see that our model explains such particles fairly well. The deviation from the straight line is due to the diquark mass.

Next we go over to the discussion on the decay width of the tetraquark. Here it is interesting to notice that our tetraquarks with $E<2$GeV can not decay with the process of the pair production by virtue of the conservation of energy. Because the total energy of the decay products (a baryon and an antibaryon) must be larger than about $2$GeV. Therefore it is predicted that the decay widths of $f_{}$(1110), $f_{0}$(1500) are suppressed and those of $f_{0}$(2020) and $f_{0}$(2200) are enhanced. The results are plotted as circles on the $(M,\Gamma)$ plane \cite{PDG} as shown in Fig.5. 
\begin{figure}
 \includegraphics[width=\linewidth]{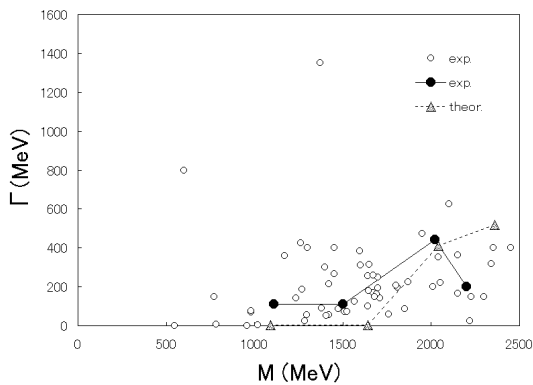}
 \caption{The masses and decay widths of all the unflavored mesons observed experimentally are plotted by the open circles. The filled ones represent the candidates for the vibrational excited states of the tetraquark. The corresponding calculated states are denoted by the triangles.}
\end{figure}
The decay mode to baryon-antibaryon has not been observed yet in the case of $f_{0}$(2020) \cite{Bea00,Bea01}. The experiments in Fig.5 is focused on the process $pp\to p_{f}(4\pi)p_{s}$. We consider that even if a baryon-antibaryon pair is created in the process, the antibaryon would interact at once the other baryon and decay into pions. As for the decay of $f_{0}$(2200), it is seen in the experiment $\bar{p}p\to \pi\pi$ where only the $\pi\pi$ mode is selected \cite{HB94}. Although it is not detected in the experiments, we conjecture a resonance reaction of $\bar{p}p\to \bar{p}p$. This resonance would support the existence of the tetraquark considered in our model.

The calculated decay widths are shown by the triangles. The black circles represent $f_{}$(1110), $f_{0}$(1500), $f_{0}$(2020), $f_{0}$(2200) and the open ones all the other unflavored mesons observed experimentally. The enhancement of the decay widths of $f_{0}$(2020), $f_{0}$(2200) may be explained by the open of the decay channel due to the pair production. The decay widths of $f_{}$(1110) and $f_{0}$(1500) are relatively small and caused by the other decay modes. Of course it is not adequate to expect quantitative discussions on the experimental values by our model. Because our model is oversimplified and the data on the decay widths have large ambiguity. Here we are satisfied with that the calculated decay widths are the same order of magnitude as the experimental data.

\section{Discussions and Summary}

In the previous section, we have studied the scalar mesons above 1GeV from the standpoint of the $qq-\bar{q}\bar{q}$ model. There are, however, the other scalar mesons not described by the present model. In fact they seem to be $f_{0}$(1370) and $f_{0}$(1710). There is a possibility that they are described by the usual $\bar{q}-q$ model. We will investigate such scalar mesons using the simple $\bar{q}-q$ model in this section.

Let us consider a meson composed of $q$ and $\bar{q}$ connected by the color flux. If the meson is $p$-wave, we have the scalar meson, because the intrinsic parity of $\bar{q}$ is minus. The mass spectrum of this system is calculated in the same manner as was done in the previous section. In particular we restrict ourselves to the case of $m=1$ ($p-wave$). The vibrational excited states are shown as a function of $n$ in Fig.6.
 \begin{figure}
 \includegraphics[width=\linewidth]{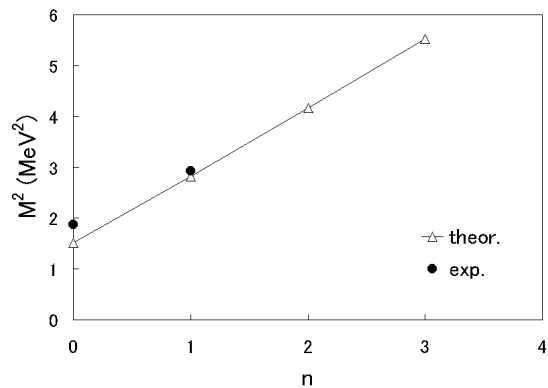}
 \caption{The vibrational excited states of the $q-\bar{q}$ mesons. The quantum number of the vibrational motion is denoted by $n$.}
 \end{figure}
The closed circles in the figure represent the corresponding particles $f_{0}$(1370) and $f_{0}$(1710) tentatively. Therefore it is suggested that these particles are composed of a quark-antiquark in the $p$-wave state. 

What is a discrimination between the $qq-\bar{q}\bar{q}$ and $\bar{q}-q$ configurations? One possible clue is the decay width. That of the four-quark particle is suppressed for the conservation of energy as mentioned before. In fact the decay width of $f_{0}$(1370) is larger than 1 GeV. However that of $f_{0}$(1710) is not so large, that is 140 MeV, which is larger than those of $f_{}$(1110) and $f_{0}$(1500) \cite{PDG}.

In conclusion, we have studied possibility of the existence of the tetraquark states which are composed of a diquark and an anti-diquark connected by a color flux. The vibrational and rotational excited states of the string are calculated with the use of the WKB approximation. It is suggested that they may be identified as the observed mesons. Moreover the decay widths of the tetraquarks are calculated using the Schwinger mechanism. They are suppressed by the conservation of energy, which is consistent with the experimental results. It is a very interesting problem to study the possibility of the tetraquark.

\begin{acknowledgments}
The authors would like to thank Prof. F.Takagi at Ishinomaki Senshu University and Prof. K.Yamawaki at Nagoya University for helpful discussions. They also thank the Yukawa Institute for Theoretical Physics at Kyoto University. Discussions during the YITP workshop YITP-W-04-07 on ``Thermal Quantum Field Theories and Their Applications" were useful to complete this work.
\end{acknowledgments}


\begin{thebibliography}{00}
\bibitem{W80}F.Wilczek, Ann.Rev.Nucl.Part.Sci. {\bf 32}, 177 (1982).
\bibitem{A01}B.Aubert, {\it et al.} (BABAR Collaboration), Phys.Rev.Lett. {\bf 90} 242001 (2003).
\bibitem{J77a}R.L.Jaffe, Phys.Rev. {\bf D15}, 267 (1977).
\bibitem{J77b}R.L.Jaffe, Phys.Rev. {\bf D15}, 281 (1977).
\bibitem{R04}D.P.Roy, J.Phys.G:Nucl.Part.Phys. {\bf30}, R113-R121 (2004).
\bibitem{Z00}A.Zhang, Phys.Rev. {\bf D61}, 114021 (2000).
\bibitem{JW03}R.L.Jaffe and F.Wilczek, Phys.rev.Lett. {\bf 91}, 232003 (2003).
\bibitem{G04}V.Gupta, hep-ph/0407289.
\bibitem{LZDL04}Y.-R.Liu, Shi-Lin Zhu, Y.-B.Dai, and C.Liu, hep-ph/0407157.
\bibitem{MPPR04}L.Maiani, F.Piccinini, A.D.Polosa, and V.Riquer, Phys.Rev.Lett. {\bf 93}, 212002 (2004).
\bibitem{MPPR05}L.Maiani, F.Piccinini, A.D.Polosa, and V.Riquer, hep-ph/0501077.\bibitem{APE93}M.Anselmino, E.Predazzi, S.Ekelin, S.Fredriksson, and D.B.Lichtenberg, Rev.Mod.Phys. {\bf 65}, 1199 (1993).
\bibitem{IP83}N.Isgur and J.Paton, Phys.Lett. {\bf 124B}, 247 (1983).
\bibitem{MKB89}A.B.Migdal, S.B.Khokhlachev and V.Yu.Borue, Phys.Lett. {\bf B228}, 167 (1989).
\bibitem{M90}A.B.Migdal, Nucl.Phys. {\bf A518}, 358 (1990).
\bibitem{IT99}M.Iwasaki and F.Takagi, Phys.Rev. {\bf D59}, 094024 (1999).
\bibitem{INST03}M.Iwasaki, S.Nawa, T.Sanada and F.Takagi, Phys.Rev. {\bf D68}, 074007 (2003).
\bibitem{JT75}K.Johnson and C.B.Thorn, Phys.Rev. {\bf D13}, 1934 (1976).
\bibitem{B01}G.S.Bali, Phys.Rep. {\bf 343}, 1 (2001).
\bibitem{AK05}C.Alexandrou and G.Koutsou, hep-lat/0407005.
\bibitem{PDG}S.Eidelman {\it et al.} (Particle Data Group), Phys.Lett.B {\bf 592}, 1 (2004).
\bibitem{S51}J.Schwinger, Phys.Rev. {\bf 82}, 664 (1951).
\bibitem{G79}E.G.Gurvich, Phys.Lett. {\bf 87B}, 386 (1979).
\bibitem{CNN79}A.Casher,H.Neuberger and S.Nussinov, Phys.Rev. {\bf D20}, 179 (1979)
\bibitem{GM83}N.K.Glendenning and T.Matsui, Phys.Rev. {\bf D28}, 2890 (1983).
\bibitem{Bea00}D.Barberis {\it et al.}, Phys.Lett.B {\bf 471}, 440 (2000).
\bibitem{Bea01}D.Barberis {\it et al.}, Phys.Lett.B {\bf 484}, 198 (2000).
\bibitem{HB94}A.Hasan and D.V.Bugg, Phys.Lett.B {\bf 334}, 215 (1994).

\end{thebibliography}
\end{document}